\documentclass[preprint,preprintnumbers,amsmath,amssymb]{revtex4}
\usepackage{graphicx}
\usepackage{epsfig}
\usepackage{lscape}
\usepackage{dcolumn}
\usepackage{bm}
\begin{document}
\title{ Time-dependent density functional theory calculation of van der Waals coefficient of sodium clusters}
\author{Arup Banerjee$^{a}$, Aparna Chakrabarti$^{b}$, and Tapan K. Ghanty$^{c}$ \\
(a) Laser Physics Application Division, Raja Ramanna Centre for Advanced Technology\\
Indore 452013, India\\
(b) Semiconductor Laser Section, Raja Ramanna Centre for Advanced Technology\\
Indore 452013, India\\
(c) Theoretical Chemistry Section, Chemistry Group, \\
Bhabha Atomic Research Centre,\\
Mumbai 400 085, India}
\begin{abstract}
In this paper we employ all-electron \textit{ab-initio} time-dependent density functional theory based method to calculate the long range dipole-dipole dispersion coefficient (van der Waals coefficient) $C_{6}$ of sodium atom clusters containing even number of atoms ranging from 2 to 20 atoms. The dispersion coefficients are obtained via Casimir-Polder relation. The calculations are carried out with two different exchange-correlation potentials: (i) the asymptotically correct statistical average of orbital potential (SAOP) and (ii) Vosko-Wilk-Nusair representation of exchange-correlation potential within local density approximation. A comparison with the other theoretical results has been performed. We also present the results for the static polarizabilities of sodium clusters and also compare them with other theoretical and experimental results. These comparisons reveal that the SAOP results for $C_{6}$ and static polarizability are quite accurate and very close to the experimental results. We examine the relationship between volume of the cluster and van der Waals coefficient and find that to a very high degree of correlation $C_{6}$ scales as square of the volume. We also present the results for van der Waals coefficient corresponding to cluster-Ar atom and cluster-N$_{2}$ molecule interactions. 
\end{abstract}
 \maketitle 
 \section{Introduction}
 The long range dispersive or van der Waals forces play a significant role in the description of many physical and chemical phenomena such as adhesion, surface tension, physical adsorption, etc. These forces originate from the correlations between electron density fluctuations at widely separated locations. The van der Waals interaction between two neutral polarizable molecules has $R^{-6}$ dependence (provided orientational averages have been performed), where $R$ is the separation between the two molecules. The van der Waals coefficient $C_{6}$ associated with the $R^{-6}$ dependent dispersive interaction describes the dipole-dipole interaction between two polarizable systems. This paper is devoted to the \textit{ab-initio} time-dependent density functional theory (TDDFT) based calculation of $C_{6}$ for sodium-cluster-cluster, sodium-cluster-argon-atom, and sodium-cluster-nirogen-molecule interactions. For our calculations we consider closed-shell sodium clusters containing up to 20 atoms.

The knowledge of van der Waals coefficient $C_{6}$ is useful for the description of cluster-cluster collisions \cite{amadon} and also for characterizing the orientation of clusters in bulk matter \cite{gunnarsson,lambin}. We note here that large body of theoretical \cite{koutecky,brackreview,deheer,haberland,alonso,ekardtrev} 
and experimental \cite{deheer,knight,brechignac,rayane,tikhonov} work on the electronic and optical response properties of sodium atom clusters exists in the literature. Majority of the theoretical calculations on sodium clusters have been performed by employing density functional theory (DFT) or its time dependent version TDDFT within the spherical jellium  background model (SJBM) (see the reviews \cite{brackreview,alonso}). The SJBM replaces the discrete ionic structure of clusters by a spherically symmetric uniform positive charge background and thus making it possible to carry out calculations for the optical response properties of reasonably large clusters of around 100 atoms \cite{brackreview,madjet}. Parallel to the jellium model calculations, several DFT and TDDFT based all-electron \textit{ab-initio} and pseudopotential calculations devoted to the ground state and the optical response properties of sodium clusters taking into account the actual geometrical arrangement of the sodium atoms  have been reported in the literature \cite{martins,moulett,andreoni,guan,calmanici,kummel,kronik,blundell,pacheco3,solovyov,ghanty}. However, these calculations could handle clusters with smaller sizes than the ones that could be studied by performing jellium based calculations.

We note here that only very few papers devoted to the calculation of the van der Waals coefficients and their measurements exist in the literature. In Refs. \cite{pacheco1,pacheco2,serra}, time-dependent Kohn-Sham (TDKS) equation of TDDFT within SJBM was employed to calculate the van der Waals coefficients. On the other hand, in Ref. \cite{banerjee}, a purely density-based modified Thomas-Fermi approach within TDDFT has been applied to calculate the coefficients.  It is only very recently that the first all-electron \textit{ab-initio} calculation of the van der Waals coefficient $C_{6}$ of small sized closed shell sodium cluster containing up to 20 atoms has been reported in the literature \cite{jiemchooroj}. In Ref. \cite{jiemchooroj} calculations have been carried out by employing linear complex polarization propagator approach in conjunction with Hartree-Fock method and TDDFT formalism with hybrid B3PW91 exchange-correlation (XC) functional \cite{becke1,perdew1}. For the calculations of response properties by employing TDDFT approach one needs to use approximate forms for the XC functionals. It has been demonstrated that the accuracy of the results for the response properties obtained via TDDFT crucially depend on the nature of the XC potential, specially its behaviour in the asymptotic region \cite{gisbergen1,banerjee1}. Keeping this in mind, we carry out all electron TDDFT based calculation of van der Waals coefficient $C_{6}$ between clusters of sodium atoms of various sizes with a XC potential possessing correct asymptotic behaviour. We employ a model potential, called statistical average of orbital potential (SAOP) which has got desirable properties both in the asymptotic and the inner regions of a molecule \cite{gritsenko,schipper}. In order to study the effect of XC potential on the results for $C_{6}$,  we make a detailed comparison of our results with the corresponding data of Ref. \cite{jiemchooroj} which were obtained with a different XC potential. Due to unavailability of any experimental data on $C_{6}$ for sodium-cluster-cluster interaction no comparison could be made with the experimental results. At this point it is important to note that Kresin and Scheidemann \cite{kresin1} measured integral scattering cross section in low energy collision experiments between a beam of sodium clusters and Ar- or N$_{2}$-vapour. For low energy collisions the integral scattering cross section depends on the van der Waals coefficient $C_{6}$. The experimental results for integral scattering cross section matched quite well with the theoretical predictions which were obtained by employing London dispersion formula  for $C_{6}$. This formula has also been employed in Ref. \cite{cole} to calculate the van der Waals coefficients corresponding to cluster-cluster interaction. The London dispersion formula is valid under single pole approximation which assumes that all the strength of the dipole transition is concentrated in a single peak. We carry out \textit{ab-initio} TDDFT based calculations of $C_{6}$ for Na$_{n}$-Ar and  Na$_{n}-$N$_{2}$ interactions (where n is an even integer lying in the range 2 to 20) with the SAOP and compare them with the results used in Ref. \cite{kresin1} to reproduce their experimental data. The main motivation for such a comparison is to establish the applicability and assess the accuracy  of London's formula in the calculation of $C_{6}$ for the above-mentioned cluster-atom and cluster-molecule systems. 
 
Moreover, it is well known that the results for static dipole polarizability of sodium atom clusters obtained by employing DFT within SJBM are generally underestimated in comparison to the corresponding experimental as well as \textit{ab-intio} data. However, such  comparisons of the results for $C_{6}$ obtained by employing jellium model and \textit{ab-initio} calculations have not been made. In order to test the accuracy of jellium model, we compare the numbers for $C_{6}$ of 2, 8, and 20 atom clusters obtained by employing LDA-XC potential in the realm of SJBM  with the corresponding \textit{ab-initio} results.
Such study is important, as the jellium model often turns out to be much more efficient tool particularly when dealing with larger cluster systems. Our study clearly reveals that like polarizability the results for $C_{6}$ obtained via jellium model are also reasonably accurate. 
 
Before proceeding further, it is important to note that
DFT in principle should give the exact
ground-state properties including the long range van der Waals energies.
However, the widely used LDA and generalized gradient approximations (GGA)
\cite{gga1,gga2,gga3} XC functionals fail to reproduce the van der Waals
energies. This is due to the fact that the LDA and the GGA cannot
completely simulate the correlated motion of electrons arising
from Coulomb interaction between distant non overlapping
electronic systems. It is only recently that attempts
\cite{andersson,dobson,kohn} have been made to obtain van der
Waals energies directly from the ground-state energy functional by
correcting the long range nature of the effective Kohn-Sham
potential. On the other hand, it is possible to make reliable
estimates of the van der Waals coefficient $C_{6}$ by using expressions
which relate this coefficient to the frequency dependent
dipole polarizabilities at imaginary frequencies
\cite{casimirpolder,stone}. We follow the latter route for the calculation of 
these coefficients.

The paper is organized as follows: In section II, we discuss the theoretical method and the expressions employed to calculate the van der Waals coefficient $C_{6}$ from the frequency dependent dipole polarizability. Results of our calculations are presented in Section III. 

\section{Theoretical Methods}
In order to calculate the van der Waals coefficient $C_{6}$, we make use of the Casimir-Polder expression which relates $C_{6}$ to the frequency dependent dipole polarizability evaluated at imaginary frequency. In accordance with this expression the orientation averaged dispersion coefficient between two molecules A  and B is given by \cite{casimirpolder,stone}
\begin{equation}
C_{6}(A,B) = \frac{3}{\pi}\int_{0}^{\infty}d\omega\bar{\alpha}_{A}(i\omega
)\bar{\alpha}_{B}(i\omega ) \label{casimirpolder}
\end{equation}
where $\bar{\alpha}_{j}(i\omega)$ is the isotropic average dipole polarizability of the j-th molecule and it is given by
\begin{equation}
\bar{\alpha}_{j}(\omega) = \frac{\alpha^{j}_{xx}(\omega) + \alpha^{j}_{yy}(\omega) + \alpha^{j}_{zz}(\omega)}{3}.
\end{equation}
In the above expression $\alpha_{xx}(\omega)$, $\alpha_{yy}(\omega)$ and $\alpha_{zz}(\omega)$ are diagonal elements of the dipole polarizability tensor. Therefore, the calculation of dispersion coefficient $C_{6}$ boils down to the determination of frequency dependent dipole polarizability tensor followed by an evaluation of the quadrature. It is for the determination of the frequency dependent polarizability we use \textit{ab-initio} TDDFT based method. In this paper this task has been accomplished by using ADF program package \cite{adf}. We refer the reader to Ref. \cite{gisbergen} for detailed description of the method adopted in this package for obtaining frequency dependent polarizabilities.

As mentioned before that a TDDFT based response property calculation requires approximating the XC functional at two different levels. The first one is the static XC potential needed to calculate the ground-state KS orbitals and their energies. The second approximation is needed to represent the XC kernel $f_{XC}({\bf r},{\bf r'},\omega)$ which determines the XC contribution to the screening of an applied field. For the XC kernel, we use reasonably accurate adiabatic local density approximation (ALDA) \cite{petersilka}. On the other hand, for the static XC potential needed to calculate the  ground-state orbitals and energies, two different choices have been made.  These are (i) the standard potential under local density approximation (LDA) as parametrized by Vosko, Wilk and Nusair \cite{vwn} and (ii) the orbital dependent SAOP which is more accurate both in the inner and asymptotic regions \cite{gritsenko,schipper}. The results obtained by these two XC potentials are compared in order to investigate the effect of XC potential on the dispersion coefficients. 

The calculations of frequency dependent polarizabilities of sodium clusters are carried out by using large Slater type orbital (STO) basis sets. It is well known that for accurate calculations of response properties it is necessary to have large basis sets with both polarization and diffuse functions. For our purpose, we have chosen all electron even tempered basis set ET-QZ3P-2DIFFUSE with two sets of diffuse functions consisting of (11s,9p,7d,3f) functions for Na, (10s,8p,5d,3f) functions for Ar and (8s,6p,4d,3f) functions for N.  The application of basis set with the diffuse functions often leads to the problem of linear dependencies. Such problem have been circumvented by removing linear combinations of functions corresponding to small eigenvalues of the overlap matrix. 
We expect that the size of the chosen basis set will make our results very close to the basis-set limit.

The Casimir-Polder integral Eq. (\ref{casimirpolder}) has been evaluated by employing thirty point Gauss-Chebyshev quadrature scheme as described in Ref. \cite{rijks}. The convergence of the results have been checked by increasing number of frequency points.

In order to perform \textit{ab-initio} calculation of response properties we need to choose the ground-state geometries of clusters. For dimer $Na_{2}$ we have used experimental bond length 3.0786 $\AA$. On the other hand, for larger clusters (4- to 20-atom clusters) we use the structures which are obtained via geometry optimization calculation with triple-$\xi$ with two added polarization functions (TZ2P basis set) and Becke-Perdew (BP86) XC potential \cite{becke2,perdew2}. All the optimizations are carried out with the convergence criteria for the norm of energy gradient and energy, fixed at $10^{-4}$atomic units and $10^{-6}$atomic units, respectively. The optimized structures obtained by us are in agreement with the corresponding results of Refs. \cite{solovyov,ghanty}. In case of a cluster having more than one isomers, we choose the one possessing the lowest energy for our calculations of the dipole polarizability. Here we note that our geometry optimization calculation for the cluster containing 20 sodium atoms yields lower energy for the structure with C$_{2v}$ symmetry than the one with T$_{d}$ symmetry. This is in contrast to the result of Ref. \cite{solovyov}.  The next section is devoted to the discussion of the results for the dispersion coefficient $C_{6}$ obtained by us.
\section{Results and Discussion}
We begin this section with the discussion on the results for $C_{6}$ between similar pair of sodium clusters (Na$_{n}$-Na$_{n}$) with even number of atoms $n$ ranging from 2 to 20. These results are shown in Fig. 1 along with the corresponding theoretical results of Ref. \cite{jiemchooroj}. For completeness, we also show in Fig. 1 the results obtained by us with the LDA XC energy functional. It can be clearly seen from Fig. 1 that the results of the SAOP are higher than the corresponding data obtained by employing the hybrid B3PW91 potential except for the 14 atom cluster case. For $Na_{14}- Na_{14}$ case the SAOP yields $C_{6} = 160.75\times 10^{3}$ atomic units which is approximately 3$\%$ lower than the hybrid B3PW91 result.  The discrepencies between the SAOP and B3PW91 results are very small for clusters up to 8 atoms. However, the mismatch between the two results grows for larger clusters. We note here that the results obtained by SAOP follow a monotonically increasing trend with the increase in the number of constituent atoms of the cluster. On the other hand, results of Ref. \cite{jiemchooroj} shows no regular trend. It can also be seen from Fig. 1 that the results for $C_{6}$ obtained with the LDA-XC potential are systematically lower than the corresponding SAOP data. This is consistent with the fact that the LDA-XC potential fails to exhibit correct behaviour both in the inner and asymptotic regions of the molecule -  which is required for accurate determination of the frequency dependent dipole polarizability. 

In order to test the accuracy of the SAOP results, we make a comparison of the results obtained with a large basis set coupled cluster model with single and double excitations (CCSD) \cite{jiemchooroj} for very small clusters like the dimer and tetramer with the corresponding SAOP values. For the dimer, SAOP and CCSD results for $C_{6}$ are 4.462$\times 10^{3}$ atomic units and 4.362$\times 10^{3}$ atomic units respectively. On the other hand, for the tetramer, SAOP results for $C_{6}$ is around 7$\%$ higher than that of CCSD calculation. 
We expect that the SAOP results is more accurate as the basis set quality in the CCSD calculation for the tetramer does not match that of dimer calculation and the discrepancy in the two results can be reduced by performing a CCSD calculation with better basis set. 
As mentioned before, the accuracy of our $C_{6}$ results obtained with SAOP can not be checked against any experimental results. However, from the comparisons with the other theoretical results, we anticipate our results are quite accurate. Encouraged by this and also for the sake of completeness, we perform calculation of $C_{6}$ for all pair of clusters (Na$_{n}$-Na$_{m}$). The results of these calculations are presented in Table I.

In the London approximation, the dispersion coefficient $C_{6}$ between two molecules is represented in terms of an effective or a characteristic frequency $\omega_{1}$ and the static polarizability $\bar{\alpha}(0)$ as 
\begin{equation}
C_{6} = \frac{3\omega_{1}}{4}\bar{\alpha}(0)^{2}
\label{londondispersion}
\end{equation}
The above expression (Eq. (\ref{londondispersion})) is obtained with the so-called single pole approximation for the frequency dependent polarizability, which assumes that one transition is dominant than the others and it alone exhausts the total oscillator strength. The London dispersion formula provides a way to correlate the van der Waals coefficient $C_{6}$ with the static polarizability.  The dispersion coefficient is proportional to the square of static polarizability. Therefore, the accuracy of $C_{6}$ crucially depends on the precision with which static polarizability is computed. Keeping this in mind we also calculate orientationally averaged static polarizability of clusters with the SAOP and LDA- XC potential. Another reason for carrying out static polarizability calculation is that unlike $C_{6}$, fairly large amount of theoretical and experimental results on the static polarizability exist in the literature, which gives us a good opportunity to test the accuracy of our results. In Fig. 2 our results for the static polarizability are illustrated and compared with the corresponding experimental and B3PW91 data of Ref. \cite{knight} and  \cite{jiemchooroj} respectively. As all the results for the even numbered clusters considered in the present paper are available in Ref. \cite{knight}, we choose to compare these experimental data with our results. 
Fig. 2 clearly shows that albeit our results obtained with the SAOP are slightly lower than the corresponding experimental results nonetheless they are quite close to the experimental results. It can also be seen from Fig. 2 that the results obtained by B3PW91 are lower than the corresponding SAOP values except for the case of $Na_{14}$. For $Na_{14}$ the SAOP result for the average polarizability is 1553.9 atomic units, on the other hand, corresponding B3PW91 value is 1596 atomic units which like $C_{6}$ is around 3$\%$ lower than the SAOP result. To further assess the accuracy of our results, we compare the SAOP results for the polarizability of the dimer and tetramer with the corresponding highly correlated CCSD values \cite{jiemchooroj}. For the dimer, the large basis set CCSD value for the polarizability is found to be 259.5 atomic units, whereas SAOP yields 265.6 atomic units. The SAOP result for the static polarizability of the tetramer is around 7$\%$ higher than the corresponding CCSD result. We note here that same order of diffferences are also observed in the $C_{6}$ results obtained by employing SAOP and B3PW91-XC potential. It then clearly demonstates that the asymtotically correct XC potential SAOP gives quite accurate results for both static polarizabilities and dispersion coffeicients of sodium atom clusters.

Before proceeding further, we note that the geometries of the clusters considered in this paper are non-spherical and consequently the polarizability tensors are anisotropic. It is then natural to investigate how the anisotropy in polarizablity evolves with the size of the cluster and also its dependence on the nature of XC potential. For this purpose, we carry out calculations of anisotropy in polarizabilty given by
\begin{equation}
|\Delta\alpha|  =  \left [\frac{3Tr{\bf\alpha}^{2} - \left (Tr{\bf\alpha}\right )^{2}}{2}\right ]^{1/2}  \qquad {\rm (general}\quad
{\rm axes})\qquad
\end{equation}
where ${\bf\alpha}$ is the second-rank polarizability tensor. The results of these calculations are displayed in Fig. 3, where anisotropy in polarizability is plotted as function of number of atoms for SAOP and LDA-XC potential. From this figure, we infer that the anisotropy in polarizability attains minimum values for magic clusters containing 2, 8 , and 20 atoms and maximum values for clusters with 4 and 14 atoms. This trend is similar for both SAOP and LDA-XC potential. These results are consistent with the fact that the magic number clusters are more symmetric than the non-magic ones. 
 
It has already been shown that the van der Waals coefficient $C_{6}$ of spherical clusters obtained within the SJBM and TDDFT varies linearly with the square of the cluster volume \cite{banerjee}. However, for clusters with non-spherical geometries, as considered in the present paper, the extension of the above mentioned relationship between $C_{6}$ and the cluster volume is not very obvious. Recently, Chandrakumar et al. \cite{ghanty} studied the relationship between the static polarizabilty and the  cluster volume of the non-spherical clusters containing 1 to 10 atoms. The volume of the clusters has been obtained from the scaled van der Waals radius of sodium atom as suggested by Tomasi and Perisco \cite{tomasi} and the study showed that the static polarizability displays a linear dependence on the cluster volume. We examine the size-to-property relationship for the van der Waals coefficient by plotting $C_{6}$ as a function of $(volume)^{2}$ along with the least square fitted line in Fig. 4a. For completeness, we also show the average static polarizability as a function of volume of the clusters in Fig. 4b. It can be clearly seen from Fig. 4a that a good fitting is obtained with the correlation coefficient value of 0.9989. This suggests that even for non-spherical sodium clusters a good correlation exists between the van Waals coefficient and cluster volume and $C_{6}$ exhibits a linear dependence on the square of the cluster volume. The linear scaling of $C_{6}$ with $(volume)^{2}$ can be explained with the help of Eq. (\ref{londondispersion}) provided the characteristic frequency $\omega_{1}$ becomes independent of the cluster size. Under this condition $C_{6}$ scales as square of the static polarizability which in turn varies linearly with the cluster volume \cite{ghanty}. We find that the characteristic frequencies determined from Eq. (\ref{londondispersion}) with the SAOP values of $C_{6}$ and $\bar{\alpha}(0)$ show a small spread ranging from 0.08 to 0.095 atomic units around the mean value of 0.089 atomic units. As $\omega_{1}$ is almost independent of the size of the cluster the linear dependence of $C_{6}$ on the $(volume)^{2}$ is satisfied to a very high degree of correlation. This linear correlation between the van der Waals coefficient and $(volume)^{2}$ is an important result as it enables us to construct a size-to-property relationship for the van der Waals coefficient $C_{6}$. This relationship can be exploited to predict the van der Waals coefficient of larger clusters for which performing all-electron \textit{ab-initio} calculations may be very expensive if not impossible. 
A very good fitting is also obtained for the static polarizability and it is illustrated in Fig. 4b. The correlation coefficient value for the polarizability fitting is 0.9974 clearly indicating that the static polarizability scales linearly with the the volume of the cluster as found in Ref. \cite{ghanty}.  
 
Now we proceed to compare the dispersion coefficient $C_{6}$ obtained by employing jellium model \cite{pacheco1,pacheco2} with the results obtained by all electron \textit{ab-initio} calculation. The jellium based results for $C_{6}$ are obtained for 2, 8, and 20 atom clusters within the KS formalism of TDDFT by using Gunnarsson-Lundquivst (GL) \cite{gl} parameterization of the LDA-XC functional. On the other hand, in the present paper the \textit{ab-initio} results for $C_{6}$ are obtained with the VWN parametrization of LDA-XC functional. Both forms for the XC functional use the same Dirac exchange energy functional but the parameterization for the correlation part is different. We expect that this deviation will be significantly smaller than the difference in the two results arising due to the consideration of structures of the clusters in \textit{ab-initio} calculations. The comparison is made in Table II. It can be seen that the results obtained within jellium model are reasonably close to the their \textit{ab-initio} counterparts but they are systematically lower than the corresponding \textit{ab-initio} values. The maximum discrepancy between the two results is observed for 2 atom cluster. As the number of atoms in the cluster increases the gap between the the jellium based result and the corresponding \textit{ab-initio} result reduces. For example, the difference between the two results for $C_{6}$ corresponding to the pair Na$_{8}$-Na$_{8}$ is of the order of $4\%$ and it reduces to just around $2\%$ for the pair Na$_{20}$-Na$_{20}$. These results clearly demonstrate that the results for the $C_{6}$ obtained within the SJBM are quite accurate and the model is suitable for larger clusters for which \textit{ab-intio} calculations may be difficult to perform.

Finally, we discuss the results obtained for $C_{6}$ corresponding to Na$_{n}$-Ar and Na$_{n}$-N$_{2}$ interactions.  As mentioned before an experiment involving measurement of integral scattering cross section from the collisions between a sodium cluster beam and Ar- or $N_{2}$-vapour had been performed by Kresin and Scheidemann \cite{kresin1}. It has been shown in Ref. \cite{kresin1} that the values of $C_{6}$ calculated from the London dispersion formula (generalization of Eq. (\ref{londondispersion}) for two different molecules)
yield results for the integral scattering cross sections which show a good agreement with the experimental data. For details on the values of dipole transition frequencies and static polarizabilities employed to calculate $C_{6}$ we refer the reader to Ref. \cite{kresin1}. In this paper, we compare London formula based numbers for $C_{6}$ with our SAOP results. In Fig. 5 and 6, we display $C_{6}$ coefficient for the pairs $Na_{n}-Ar$ and  $Na_{n}-N_{2}$ as functions of number of atoms present in the cluster, respectively. The match between the two results both for Ar atom and N$_{2}$ molecule are quite good even though the London's formula is valid under single pole approximation and also it does not take anisotropic nature of the clusters into account. These results indicate that the approximate London dispersion formula is well suited for calculating dispersion coefficient $C_{6}$ for the $Na_{n}-Ar$ and  $Na_{n}-N_{2}$ interactions.  

\section{Conclusion}
The van der Waals coefficient $C_{6}$ for the sodium atom clusters containing even number of atoms ranging from 2 to 20 atoms have been calculated by employing all-electron \textit{ab-initio} method within the realm of TDDFT. The calculations are performed by using a model XC potential SAOP having correct behaviour both in the asymptotic and inner regions of the molecule.   The van der Waals coefficient is obtained by using Casimir-Polder expression which needs frequency dependent dipole polarizabilties of the two interacting species. All the calculations are carried out with one of the largest STO basis sets. In this paper the performance of the SAOP for the calculations of the static polarizability and van der Waals coefficient, $C_{6}$ of sodium clusters has been investigated against other available theoretical and experimental results. We find that the SAOP results for the static polarizabilities are  closer to the experimental data than other theoretical results. There are no experimental results available in the literature for the van der Waals coefficient of sodium clusters. However, the dependence of $C_{6}$ on the polarizabilities of two interacting species and the SAOP results for the polarizabilities suggest that the SAOP results for the $C_{6}$ must be quite accurate. The van der Waals coefficient between same pair of clusters obtained by SAOP is found to have good linear correlation with the square of the cluster volume. This scaling law can be exploited to determine $C_{6}$ of larger clusters. Moreover, in accordance with the earlier studies, we also find a very good linear correlation between the static polarizability and the volume of clusters. The performance of the SAOP is also examined by calculating van der Waals coefficients for the pairs $Na_{n}$-Ar and $Na_{n}$-$N_{2}$. These results are compared with the ones obtained from the London's formula which were used to fit the experimental data of the scattering cross section for the cluster-atom and cluster-molecule collisions. We find a close match between the two results for both cluster-atom and cluster-molecule cases. 
In this paper we also carry out a systematic assessment of the accuracy of the jellium based calculation of $C_{6}$ by comparing the results obtained within the SJBM with the corresponding all-electron \textit{ab-intio} values. We conclude from this comparison that jellium based results for $C_{6}$ are reasonably accurate and the jellium model becomes increasingly more suitable for larger clusters. 

\acknowledgments{ We wish to thank Mr. Pranabesh Thander of RRCAT Computer Centre for his help and support in providing us the uninterrupted computational resources and also for smooth running of the codes. It is a pleasure to thank Prof. Vitaly Kresin for his valuable suggestions and making his numbers for $C_{6}$ available to us. Thanks are also due to Dr. S. J. A. van Gisbergen for his help.}

 .

\clearpage
\newpage
\section*{Figure captions}
{\bf Fig.1} Plot of van der Waals coefficient $C_{6}$ of sodium atom clusters in atomic units obtained with different XC potentials. The lines joining the points are guide to the eye. The B3PW91 results are taken from Ref. \cite{jiemchooroj}. 
 
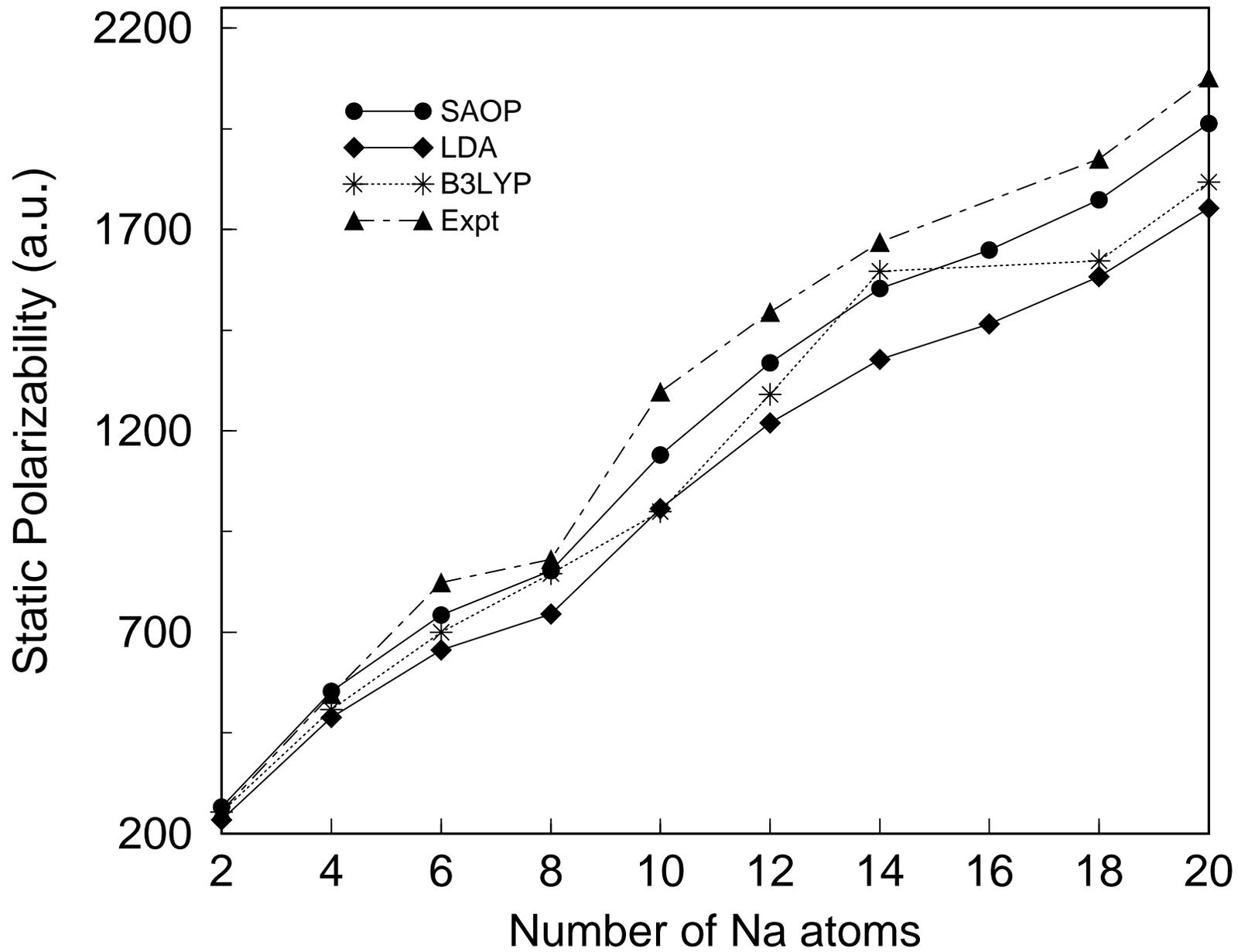
{\bf Fig.2} Plot of average static polarizability $\bar{\alpha}(0)$ of sodium atom clusters in atomic units obtained with different XC potentials along with the experimental data. The lines joining the points are guide to the eye. The B3PW91 results are taken from Ref. \cite{jiemchooroj} and the experimental data are from Ref. \cite{knight}.

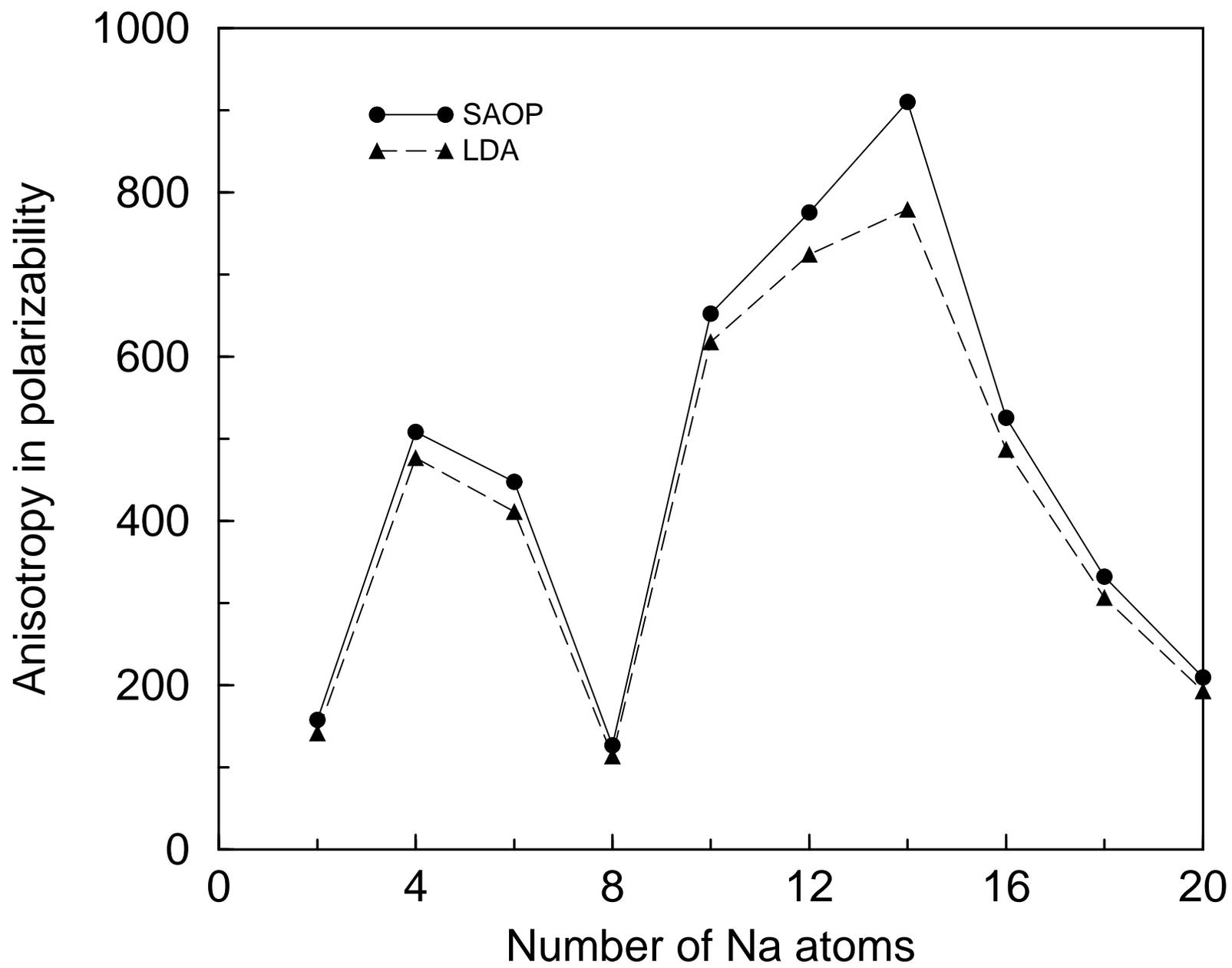
{\bf Fig.3} Plot of anisotropy in polarizability $\Delta{\alpha}$ of sodium atom clusters in atomic units obtained with different XC potentials. The lines joining the points are guide to the eye. 

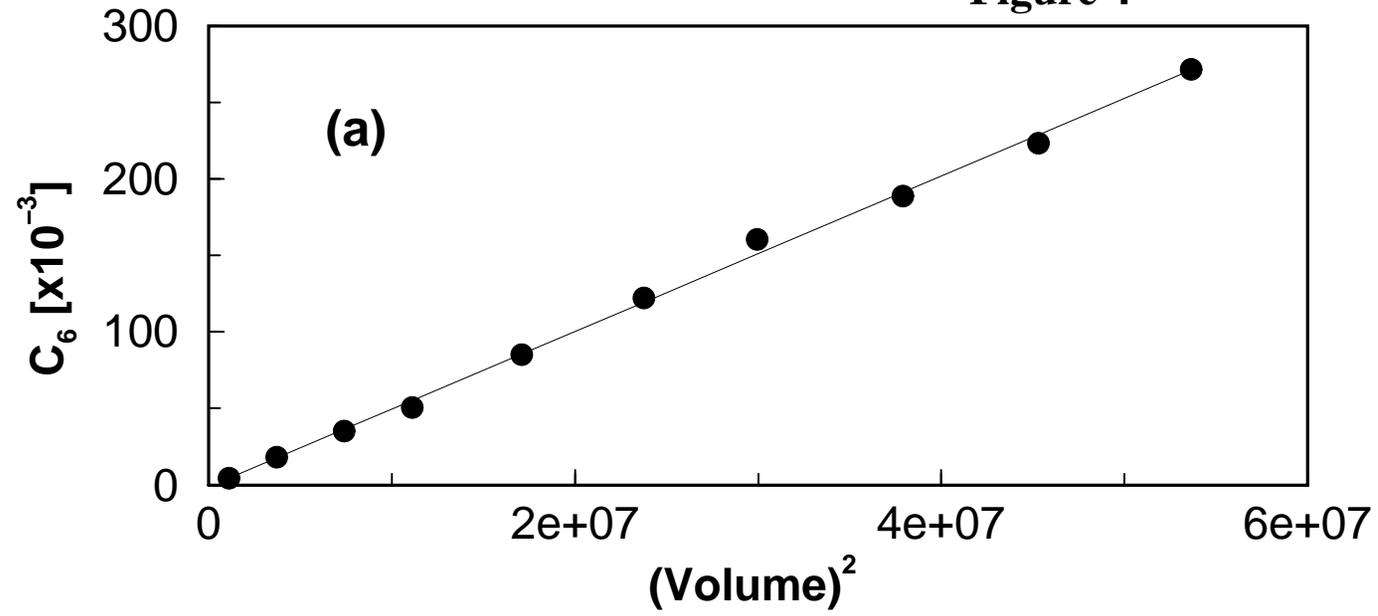
{\bf Fig.4} Plot of (a) van der Waals coefficient $C_{6}$ and (b) average static polarizability obtained with SAOP as functions of square of the cluster volume and cluster volume respectively.  All the results are in atomic units and straight lines are least square fitted lines.

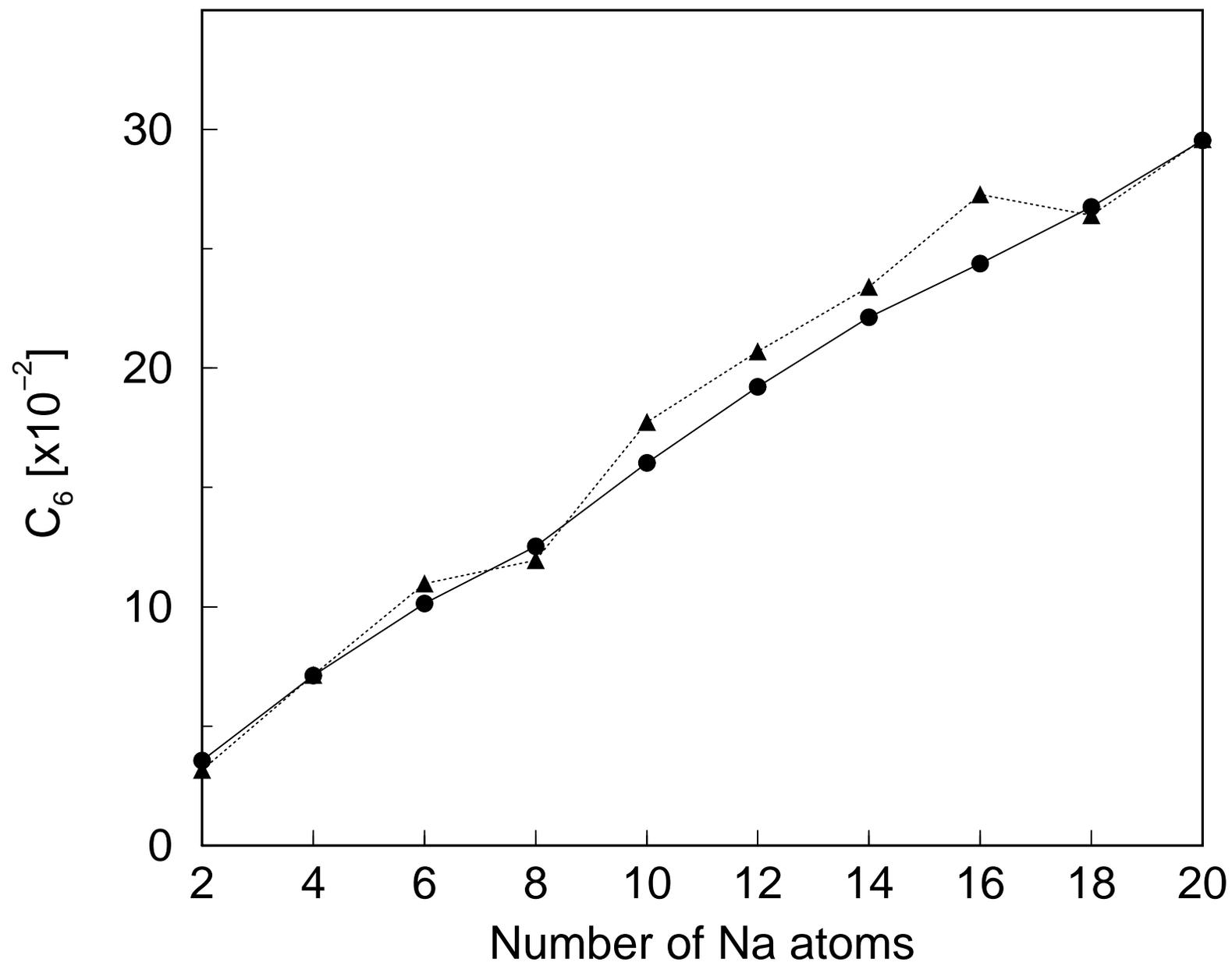
{\bf Fig.5} Comparison of all-electron \textit{ab-intitio} ( solid circle) and London formula based (solid triangle) results for the van der Waals coefficient $C_{6}$ ($\times 10^{-2}$) corresponding to the pair $Na_{n}$-Ar. The \textit{ab-initio} results are obtained with the SAOP and all the results for $C_{6}$ are in atomic units. The lines joining the points are guide to the eye..

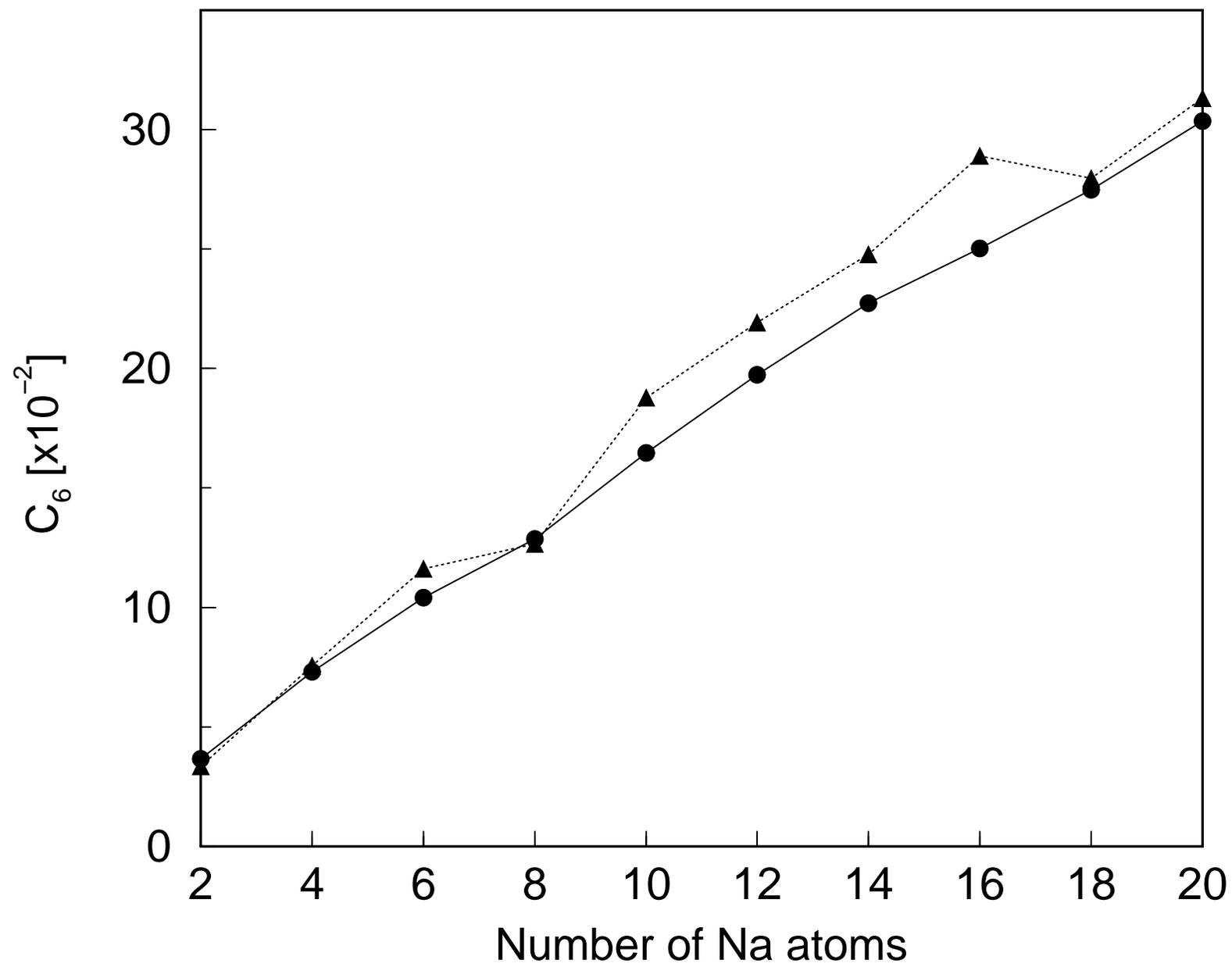
{\bf Fig.6} same as Fig. 5 but for the pair $Na_{n}$-$N_{2}$.
\newpage
\begin{table}
\caption{The dispersion coefficient $C_{6}$ ($\times 10^{-3})$ between sodium clusters in atomic units obtained with the SAOP} 
\tabcolsep=0.15in
\begin{center}
\begin{tabular}{ccccccccccc}
N  & 2 & 4 & 6 & 8 & 10 & 12 & 14 & 16 & 18 & 20 \\
\hline
2 & 4.46 & 9.00 & 12.55 & 15.00 & 19.47 & 23.35 & 26.76 & 28.99& 31.52& 34.73  \\
4 &  & 18.17 & 25.31& 30.24 & 39.26& 47.08 & 53.96 & 58.42& 63.52&69.93 \\
6  &    &    & 35.30& 42.22& 54.78 & 65.70 & 75.31& 81.58 & 88.73& 97.77  \\
8  &    &    & & 50.58& 65.57 & 78.63 & 90.16& 97.75 & 106.36& 117.21  \\
10  &    &    & & & 85.04 & 101.98 & 116.92& 126.71 & 137.83& 151.89  \\
12  &    &    & & &  & 122.30 & 140.21& 151.95 & 165.29& 182.14  \\
14  &    &    & & &  &  & 160.75& 174.23 & 189.54& 208.87 \\
16  &    &    & & & &  & & 188.93 & 205.58& 226.55 \\
18 &    &    & & & &  & &  & 223.72 & 246.55 \\
20 &    &    & & &  &  & &  & & 271.71  \\
\end{tabular}
\end{center}
\end{table}

\begin{table}
\caption{ Comparison of all-electron ab-intitio and jellium based results for the van der Waals coefficient $C_{6}$ ($\times 10^{-3}$) of sodium clusters in atomic units. Both the results are obtained with LDA-XC potential. The numbers in parenthesis are results of
Refs. \cite{pacheco1,pacheco2} } \tabcolsep=0.1in
\begin{center}
\begin{tabular}{cccc}
N  & 2 & 8 & 20 \\
\hline
2 & 3.68 & 12.39 & 29.22   \\
   &(2.62)& (10.22) & (24.45)  \\
8 &  & 41.82 & 98.66 \\
   &  & (40.06) &(95.55) \\
20  &    &    & 232.81  \\
    &     &    & (228.58 )   \\
\end{tabular}
\end{center}
\end{table}

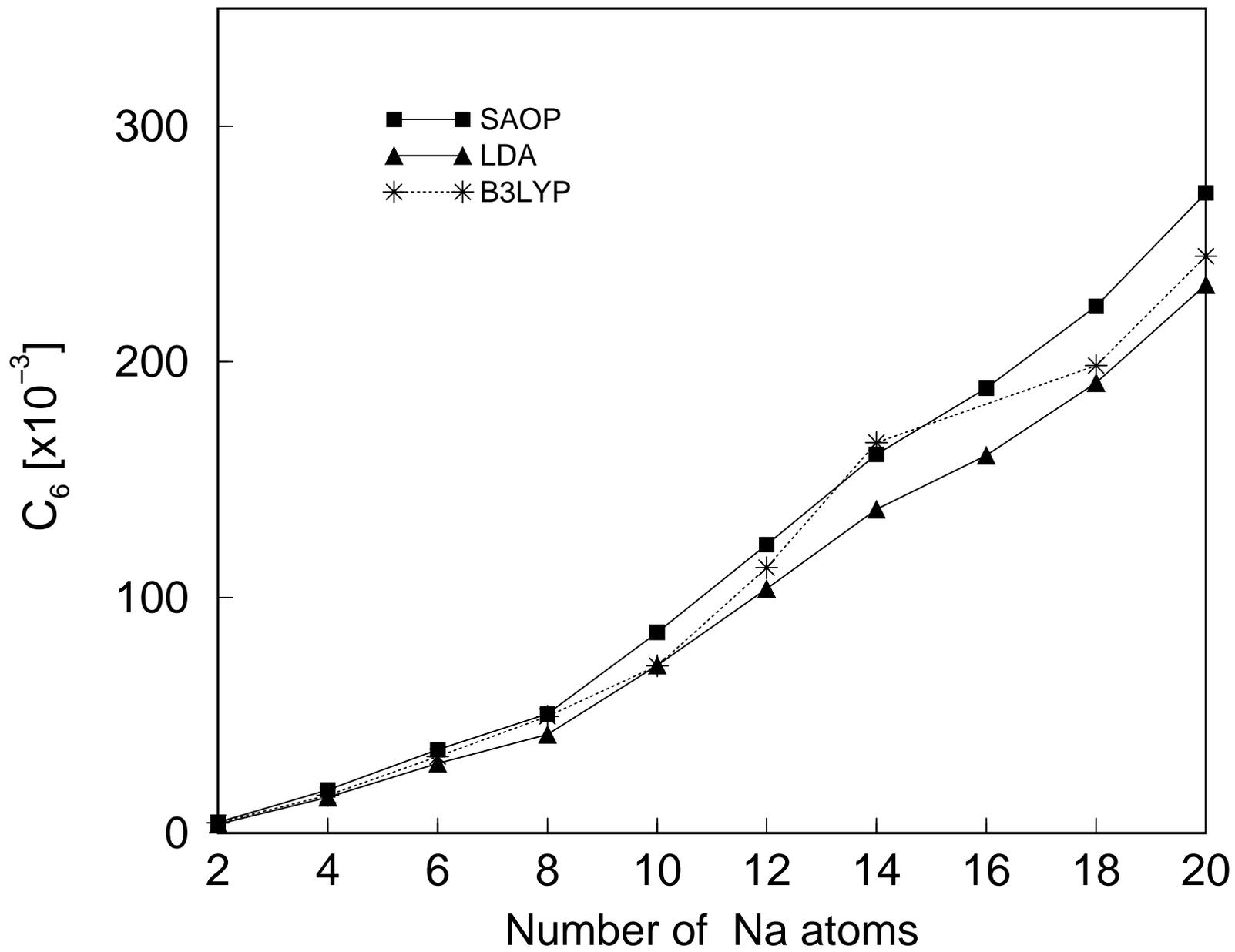
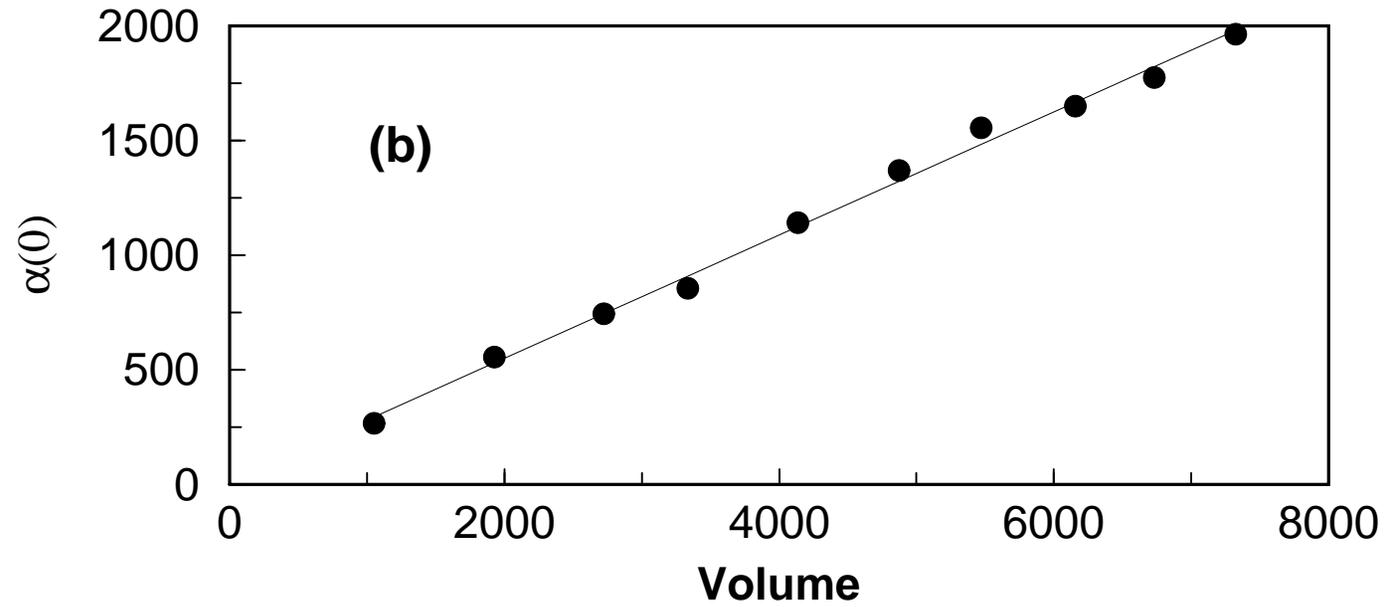
\end{document}